\documentstyle[12pt]{article}
\fussy
\begin{document}

%\title
\begin{center}
{ \Large \bf   ON ADELIC STRINGS}
\bigskip
%\author

{\large  Branko DRAGOVICH\footnote{e-mail: dragovic@phy.bg.ac.yu}}

%\address%
\date{}
\smallskip

{\it Institute of Physics, P.O.Box 57, 11001 Belgrade, Yugoslavia}  

{\it Steklov Mathematical Institute, Moscow, Russia}
\end{center}


%\maketitle%
%\abstracts%

\begin{abstract}
{New approach to $p$-adic and adelic strings,
which takes into account that not only world sheet but also
Minkowski space-time and string momenta can be $p$-adic and
adelic, is formulated. $p$-Adic and adelic string amplitudes 
are considered within Feynman's path integral formalism. 
The adelic Veneziano amplitude is calculated. Some discreteness of
string momenta is obtained. Also, adelic coupling constant is
equal to unity.} 
\end{abstract}

%\section{Introduction}
%\subsection{General Information}
%\label{subsec:general} 

{\bf 1.} Since ancient times, scientists have been interested in 
unification of human knowledge on Nature. Superstring
theory, recently resulted in the M-theory, is the best 
modern candidate  for "theory of everything". It unifies
all fundamental interactions. Gravity, nonabelian gauge
symmetry, supersymmetry and space-time dimension are
the most significant predictions of string theory. Superstring
theory is a synthesis of deep physical principles and modern 
mathematical methods. In spite of many significant achievements,
there are a number of open questions. One of the main actual 
problems is the space-time structure at the Planck scale.
To find an appropriate answer to this problem it seems unavoidable 
to extend the Riemannian geometry, which is the Archimedean one
and related to real numbers, with the non-Archimedean geometry
described by $p$-adic numbers. The corresponding mathematical
instrument suitable for unification of Archimedean and non-Archimedean 
geometries, as well as of real and $p$-adic numbers, is the space
of adeles.

Recall that the field of rational numbers $Q$ plays an important 
role not only 
in mathematics but also in physics. All numerical results of measurements 
belong to $Q$. On $Q$ there is the usual  ($\mid
\cdot \mid_\infty$) and $p$-adic ($\mid \cdot \mid_p$) absolute
value, where $p$ denotes a prime number. Completion of $Q$ with respect
to $\mid \cdot \mid_\infty$ and $\mid \cdot \mid_p$ yields the
field of real $(R\equiv Q_\infty)$ and $p$-adic $(Q_p)$ numbers, respectively. 
Many properties of $p$-adic numbers and their functions \cite{schikhof} are
very different in comparison with real numbers and classical
analysis. Any $x\in Q_p$ can be presented in the form
$$
  x= p^{\nu} (x_0 + x_1p + x_2p^2 +\cdots), \ \ x_0 \neq 0 ,
\ \  \nu\in Z ,              \eqno(1)        
$$
where $ x_i \in \{0,1,... ,p-1   \}$.

An adele $a$ \cite{gelfand} is an infinite sequence
$$
  a = (a_\infty , a_2 , a_3 , ... , a_p ,... ) ,  \eqno(2)
$$
where $a_\infty \in R$ and $a_p\in Q_p$  with restriction 
that $a_p \in Z_p = \{x\in Q_p : x= x_0 +x_1p + x_2p^2 +\cdots  \}$ 
for all but a finite
set $S$ of primes $p$. The set of all adeles $A$ may
be given in the form
$$
A =\bigcup_S {\cal A}(S) ,\ \ {\cal A}(S) = R\times \prod_{p\in S}
Q_p \times \prod_{p\notin S} Z_p .                            \eqno(3)
$$
$A$ has the structure of a topological ring.

\smallskip
{\bf 2.} Since 1987, $p$-adic numbers and adeles have been
successfully used in many branches of theoretical
and mathematical physics (for a review, see, e.g. 
\cite{freund1}, \cite{vvz}, and \cite{khrennikov}).
First applications were just in string theory. It was
suggested \cite{volovich} a hypothesis on the possible emergence
of a non-Archimedean geometry at the Planck scale
and initiated the corresponding string theory over $Q_p$
and the Galois fields. It was also proposed generalization of 
the usual Veneziano and the Virasoro-Shapiro amplitude using 
multiplicative characters over various number fields.
In \cite{freund2}, the Veneziano and the Virasoro-Shapiro
$p$-adic string amplitudes are
postulated as the Gel'fand-Graev beta functions, which
are convolutions of the real-valued characters on
the $p$-adic fields. After that it was put forward
an interesting formula \cite{freund3}: the product of the
Veneziano amplitude and its $p$-adic counterparts is
equal to unity. In other words, the Veneziano amplitude can 
be presented as the infinite product of inverses of
$p$-adic amplitudes. The similar property holds for the
Virasoro-Shapiro amplitude. The product of an ordinary string
amplitude and the corresponding $p$-adic counterparts
is usually called adelic string amplitude. However,
such adelic string amplitudes must be regularized \cite{arefeva}
and they are not equal to a constant in a more complex case.
This situation has been an obstacle to further progress in
application of $p$-adic numbers in string theory. It seems to 
be a consequence of restricted employment of $p$-adic numbers
in string amplitudes. In fact, a general feature of all these
investigations was that only world sheet may be treated $p$-adically,
while target space-time and momenta should be described by real numbers.

\smallskip
{\bf 3.} According to successful formulation of $p$-adic \cite{vv} and
adelic \cite{dragovic} quantum mechanics, approach to $p$-adic
and adelic string amplitudes should be modified. Here we give
a new proposal to $p$-adic and adelic string amplitudes based
on the following general assumptions:

{\it (i)}  space-time and matter are adelic at the Planck (string)
scale,

{\it (ii)} the Feynman path integral is an essential constituent
of quantum theory,

{\it (iii)} ordinary classical and quantum theories are effective
limits of the adelic ones.

Thus, a string is adelic, and has real and all  $p$-adic  faces
simultaneously. We use term real ($p$-adic) string when we consider
as dominant its real ($p$-adic) characteristics. In addition to
the adelic Minkowski space-time, string world sheet and 
momenta are also adelic.
However, string amplitude remains complex-valued in all these cases.

Recall that quantum amplitudes defined by means of path integral
may be symbolically presented as
$$
  A(K) = \int A(X) \chi \left(-\frac{1}{h} S[X]\right) {\cal D}X ,  \eqno(4)
$$
where $K$ and $X$ denote  classical momenta and  
configuration space, respectively. $\chi (a)$ is an additive character,
$S[X]$ is a classical action and $h$ is the Planck constant.

\smallskip
{\bf 4.} In the sequal, we briefly consider simple $p$-adic and 
adelic bosonic string amplitudes based on the functional integral
(4). As is known \cite{green}, the scattering of two real bosonic 
strings in 26-dimensional space-time at the tree level can be described 
in terms of the path integral in 2-dimensional quantum field theory
formalism as follows:
$$
  A_\infty (k_1,...,k_4) = g^2_\infty \int {\cal D}X
  \exp\left(\frac{2\pi i}{h}S_0[X]\right) 
$$
$$
\times\prod_{j=1}^4 \int d^2\sigma_j
  \exp{\left(\frac{2\pi i}{h} k_\mu^{(j)} X^\mu (\sigma_j,\tau_j)\right)
} , \eqno(5)
$$
where ${\cal D}X = {\cal D}X^0(\sigma,\tau) {\cal D}X^1(\sigma, \tau)
...{\cal D}X^{25}(\sigma,\tau)$, $\ \ d^2\sigma_j = d\sigma_j d\tau_j$
and
$$
  S_0[X] = -\frac{T}{2} \int d^2\sigma \partial_\alpha X^\mu
  \partial^\alpha X_\mu                                    \eqno(6)
$$
with $\alpha = 0,1$ and $\mu = 0,1,...,25$. Using the usual procedure
\cite{green} one can obtain the crossing symmetric Veneziano amplitude
$$
  A_\infty (k_1,...,k_4) = g_\infty^2 \int_R \mid x \mid_\infty^{k_1 k_2}
  \mid 1-x \mid_\infty^{k_2k_3} dx                         \eqno(7)
$$
and similarly the Virasoro-Shapiro one for closed bosonic strings.

As $p$-adic Veneziano amplitude, it was postulated \cite{freund2}
$p$-adic analogue of (7), i.e.
$$
  A_p (k_1,...,k_4) = g_p^2 \int_{Q_p} \mid x\mid_p^{k_1k_2}
  \mid 1-x\mid_p^{k_2k_3} dx ,                       \eqno(8)
$$
where only the string world sheet (parametrized by $x$) is $p$-adic.
Expressions (7) and (8) are Gel'fand-Graev beta functions on $R$
and $Q_p$, respectively. However, (8) cannot be derived from $p$-adic
analogue of (5) and one may doubt  validity of (8) as a $p$-adic
string amplitude. Therefore, we take $p$-adic analogue of (5), i.e.
$$
  A_p (k_1,...,k_4) = g_p^2 \int {\cal D}X \chi_p \left(-\frac{1}{h}
S_0[X]\right) 
$$
$$
\times\prod_{j=1}^4 \int d^2\sigma_j \chi_p \left(-\frac{1}{h}
k_\mu^{(j)}X^\mu (\sigma_j,\tau_j)\right) ,                   \eqno(9)
$$
to be $p$-adic string amplitude, where $\chi_p(u) = \exp(2\pi i 
\{ u\}_p)$ is $p$-adic additive character and $\{u\}_p$ is the
fractional part of $u\in Q_p$. In (9), all space-time coordinates
$X_\mu$, momenta $k_i$ and world sheet $(\sigma,\tau)$ are
$p$-adic.

Evaluation of (9), in an analogous way to the real case, leads to
$$
  A_p(k_1,...,k_4) = g_p^2 \prod_{j=1}^4 \int d^2\sigma_j
$$  
$$
\times\chi_p \left(\frac{\sqrt{-1}}{2hT} \sum_{i<j} k_ik_j 
\log((\sigma_i -\sigma_j)^2 + (\tau_i -\tau_j)^2)\right) .    \eqno(10)
$$
Since $p$-adic character $\chi_p$ is complex-valued and
logarithmic function is $p$-adic valued, amplitude (10) cannot be
reduced to (8). $p$-Adic logarithmic function is
$$
 \log x = \sum_{n=1}^\infty (-1)^{n+1} \frac{(x-1)^n}{n} ,
\ \  \mid x-1 \mid_p <1 .                           \eqno(11)
$$
Due to (10) and (11) one obtains the system of equations
$$
  (\sigma_i - \sigma_j)^2 + (\tau_i - \tau_j)^2 \in 1+pZ_p ,
 \ \  (i<j), \ \ i,j =1,...,4 ,                   \eqno(12)
$$
which determines possible $p$-adic values of the world sheet
parameters $\sigma_i$ and $\tau_i$.
Since (10) can be rewritten as 
$$
 A_p (k_1,...,k_4) = g_p^2 \prod_{j=1}^4 \int d^2\sigma_j
\chi_p (\sqrt{-1} 
$$
$$
\times\log\prod_{i<j}((\sigma_i - \sigma_j)^2
+ (\tau_i -\tau_j)^2)^{\frac{k_ik_j}{2hT}})       \eqno(13)
$$
it has to be also satisfied
$$
 \prod_{i<j} ((\sigma_i - \sigma_j)^2 + (\tau_i -\tau_j)^2
)^{\frac{k_ik_j}{2hT}} \in 1 + pZ_p .     \eqno(14)
$$
Combining (12) and (14) one gets
$$
 \prod_{i<j} (1 + pZ_p)^{\frac{k_ik_j}{2hT}} \in 1 + pZ_p ,
                                                  \eqno(15)
$$
what is satisfied if
$$
    \mid \frac{k_ik_j}{2hT} \mid_p \leq 1 .        \eqno(16)
$$

Since in experiments $k_i,k_j \in Q$ and (17) has place
for every $p$, it follows that string momenta have discrete
values in units $h=T=1 $.

Amplitude (14) becomes
$$
 A_p (k_1,...,k_4) = g_p^2 \prod_{j=1}^4 \int d^2\sigma_j ,
                                                \eqno(17)
$$
because $\chi_p (\sqrt{-1}\log u) = 1$.

\smallskip
{\bf 5.}  Adelic string amplitude is product of real and all $p$-adic 
amplitudes, i.e. 
$$
 A_A(k_1,...,k_4) = A_\infty (k_1,...,k_4) \prod_p 
 A_p (k_1,...,k_4) .                         \eqno(18)
$$
In the case of the Veneziano amplitude and
$(\sigma_i, \tau_j) \in {\cal A}(S) \times {\cal A}(S)$,
where ${\cal A}(S)$ is defined in (3), we have
$$
  A_A(k_1,...,k_4) = g^2_\infty \int_R \mid x \mid_\infty^{k_1k_2}
  \mid 1-x \mid_\infty^{k_2k_3} dx 
$$
$$
\times \prod_{p\in S} g_p^2
  \prod_{j=1}^4 \int d^2\sigma_j \times \prod_{p\notin S} g_p^2 .
                                                      \eqno(19)
$$

There is a sense to take adelic coupling constant as
$$
  g_A^2 = \mid g\mid^2_\infty \prod_p \mid g \mid_p ^2 = 1 ,
 \ \ 0\neq g\in Q .                                   \eqno(20)
$$

At the end, it follows that $p$-adic effects in the adelic 
Veneziano amplitude 
induce discreteness of string momenta and contribute to an
effective coupling constant in the form
$$
  g_{ef}^2 = g_A^2 \prod_{p\in S} \prod_{j=1}^4 \int d^2\sigma_j
  \geq 1 .
$$

\section*{Acknowledgments} The author wishes to thank A.I.
Studenikin and the Organizing
Committee of the 9th Lomonosov Conference on Elementary Particle
Physics for invitation to participate and give a talk.

%\section*{References}

\end{document}